# EEG CORRELATES OF VERBAL AND PERCEPTUAL REASONING

# Gwyneth Wesley Rolph

**Student ID**: W15651822
**Module Code**: 6PSYC002W
**Supervisor's Name**: Catherine Loveday

This report is submitted in partial fulfilment of the requirements of the University of Westminster for the award of BSc (Hons) Cognitive & Clinical Neuroscience.

This project is entirely my own work.

Signed: GWRolph -----------

Date: 9 April 2018-------------------------------------

# ACKNOWLEDGEMENTS

I would like to thank my tutor Catherine Loveday for helping me with my chosen research project, on a topic that is admittedly somewhat specialised but, for me, represents the culmination of a lifelong interest in human intelligence, particularly at the highest extremes, and what underlies it. Thank you so much for humouring my “special interest”. I would like to thank the staff at Westminster for offering a student from a nonconventional educational background the opportunity to do this degree, and for their support and encouragement over the last three years. A shout out to my classmates: you have been a wonderful group to work with. I would also like to thank all my participants, especially those who quite voluntarily travelled some distance in order to take part. Special thanks to Tony Steffert for loaning me his EEG equipment and helping me record and clean the data. And of course, to my family, for supporting me in my decision to go back to school in my forties to pursue my dream of doing scientific research.

# CONTENTS

# EEG CORRELATES OF VERBAL AND PERCEPTUAL REASONING

## ABSTRACT

Identifying the neural underpinnings of intelligence is one of the major goals of modern intelligence research. However, few imaging studies explore the high extremes of intelligence, and most use only a single composite IQ score. This study used quantitative EEG recordings to investigate correlations between the EEG variables of individual alpha peak frequency, relative power, and coherence, and scores on verbal and perceptual reasoning tests in participants with IQs from the low-average to exceptionally gifted ranges. Overall, the results did not reveal any pattern of significant relationships between these EEG measures and the two psychometric tests. Since this study had only a small sample, heavily skewed towards the high-IQ end of the scale, no definitive conclusions about the EEG measures and verbal and perceptual reasoning can be drawn. Nevertheless, it reveals many opportunities for further research, and suggestions are made for progressing this topic in the future.

# INTRODUCTION

The psychometric measurement of intelligence has been hailed as possibly the greatest success of psychology (Hunt, 2011), and intelligence testing reliably predicts success in education, work performance, and the demands of day-to-day living (Gottfredson, 1997). However, while psychometrics has over 100 years of research supporting its validity and reliability, for cognitive neuroscience, explaining how intelligence arises in the brain presents a complex challenge (Geake, 2005). Since Spearman (1904) identified the "positive manifold" between different cognitive tasks that led to the postulation of the g-factor, some have questioned whether psychometric g maps onto corresponding common neural substrates, or a "neuro-g" (Haier, 2009), or whether neuro-intelligence is non-unitary.

Modern neuroimaging and EEG are beginning to answer this question, but before the current study little research to date has explored first or second order intelligence factors and no known EEG research has studied IQ up to the exceptionally gifted (4SD+) range. To provide context for this study, the following topics are explored in turn: physiological measurements to assess intelligence, key structural/functional research on high intellectual ability (HIA), up-to-date EEG/IQ research, and issues regarding high-range psychometrics.

### *Early physiological ability measures*

Galton is recognised for one of the first systematic attempts to measure mental faculties through physical tasks in the late nineteenth century (Jensen, 1980). Thousands of volunteers undertook measures of, for example, sensory discrimination, motor speed, and grip strength. Findings, however, were disappointing: they correlated poorly with any intuitive notion of "intelligence", or even among themselves. Galton possessed neither the sophisticated equipment nor the statistical techniques to develop better tests. However, Galton's measurements laid the foundations for much future research using elementary tasks to measure cognitive abilities.

It was hoped that the information processing underlying intelligence could be linked to mental speed and neural efficiency, and measured using paradigms such as simple and choice reaction time, neural conduction velocity, latencies of event-related potentials, and inspection time (Deary, Austin & Caryl, 2000). If successful, these paradigms offered several advantages, such as objectivity, low participant anxiety, and cost. Unfortunately, mental chronometry has thus far over-promised. Over 20% of participants' response patterns do not fit the Hick Paradigm. Among the remainder, simple and choice reaction times only correlate approximately -.3 with intelligence (Kline, 1991), probably due to inconsistencies across the ability range: high ability participants are not substantially faster; rather, they make fewer slow responses that lower their mean scores (Mackintosh, 1998). Inspection time measures were devised to measure stimulus presentation time vis-à-vis response time, but again only showed modest correlations with intelligence (Kline, 1991) and is susceptible to practice effects, participant strategy, and attention and vigilance levels (Mackintosh, 1998). Colom et al. (2002) suggested there may have been range restriction effects if using students, since the tasks lack discriminatory power at higher ability levels. Nevertheless, there remains no consensus regarding the processes measured or their appropriate level of analysis (Deary, Penke & Johnson, 2010), and the factor structure of these elementary tasks has not been identified (Schubert et al., 2015). This research was largely abandoned as neuroimaging became increasingly available.

*Neuroimaging and intelligence*

Using neuroimaging to predict IQ is one of the major goals of modern intelligence research (Haier, 2016) and dates to the EEG research of the 1960s. Such research was predicated on little more than the premise that since, neuroelectric signals and intelligence both originated in the brain, their measurements might correlate. Deary and Caryl's 1997 review found little replication, while methodological inconsistencies made interpretation difficult. By the late 1980s PET was more available and largely superseded EEG in imaging work. Early PET intelligence research was important, however, because the focus was shifted from the controversy then surrounding psychometrics to the perceived objectivity of neuroscience (Haier, 2016). These studies, rather than locating an intelligence region, instead suggested distributed networks were involved. Duncan et al. (2000) identified a frontal lobe network which they posited as the neural substrate of g. Although the study had methodological shortcomings, its importance was firstly linking the lateral PFC to fluid reasoning, and secondly raising the profile of the neuroscientific study of intelligence. After around 2000, with increasing availability of fMRI, more fine-grained neuroscience investigation became possible.

Using neuroimaging for intelligence research has also justifiably received criticism. Participant strategies may alter cognitive loadings of tasks, with good problem-solvers structuring problems and ordering their thoughts differently, rendering different brain activation patterns (Kaufman, 2013). Evidence suggests high-ability participants flexibly alter their approach (e.g. linguistic, visuo-spatial) depending on problem type (Kang et al., 2017), meaning two individuals could obtain similar g scores but activate different neural pathways. Doppelmayr et al. (2005) recognised potential difficulties in using the same problem-solving task with participants of differing IQs: high-ability participants might utilise minimal neural resources, while low-ability participants quit in frustration. Avoidance of such confounds requires careful experimental designs.

Despite methodological challenges, sufficient high-quality neuroscience-based intelligence research exists to draw some initial conclusions. Following discussions among colleagues regarding the current state of intelligence research, Jung and Haier (2007) reviewed 37 structural and functional neuroimaging studies, and concluded that a discrete network of brain regions underpins general cognitive performance across a range of tasks. Based on these findings, the authors proposed a new theory of intelligence rooted in neuroscience: the Parieto-Frontal Integration Theory (P-FIT) of Intelligence.

*The Parieto-Frontal Integration Theory*

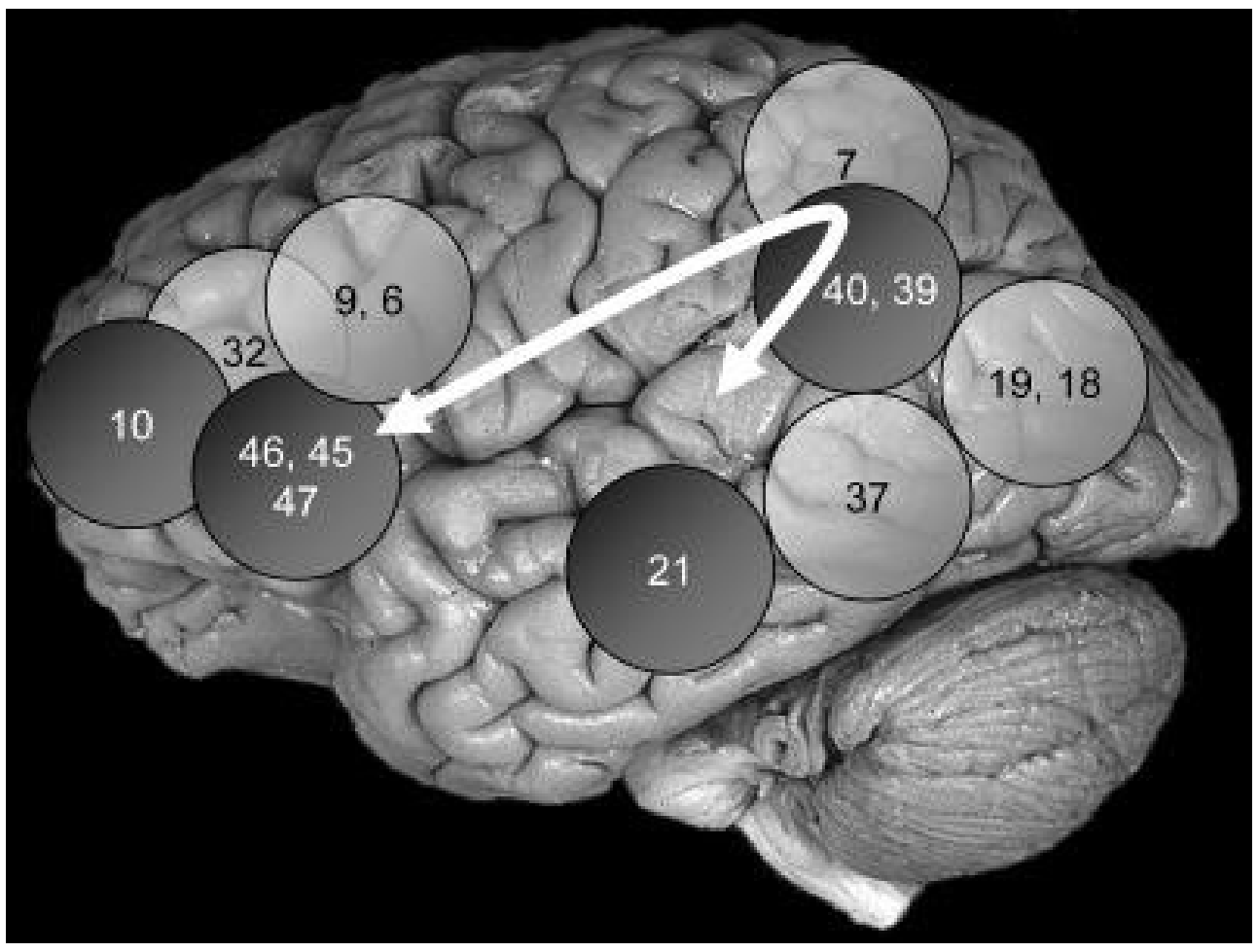

**Figure 1: Brodmann's Areas implicated in the parieto-frontal integration theory of intelligence (Image source: www.richardhaier.com)**

Under the P-FIT model, intelligence-related information processing occurs as follows: (1) Cognitively salient information is perceived and first-stage processing of visual and/or auditory information occurs in occipital and temporal cortices. Recognition and elaboration of visual imagery occurs in extrastriate cortex (Brodmann's Areas 18/19) and fusiform gyrus (BA37) and syntax of auditory information is processed in Wernicke's Area (BA22). (2) Sensory information is elaborated and abstracted in the angular gyrus (BA39), supramarginal gyrus (BA40) and superior parietal lobule (BA7). (3) Information feeds forward to frontal regions (BA6, 9, 10, 45, 46 and 47) for evaluation and decision-making. (4) The anterior cingulate cortex (BA32) is recruited in response selection and inhibiting competing responses. (5) Information communication between these processing units is assumed to be critically dependent upon the integrity of connecting white matter tracts (e.g. arcuate fasciculus). Brain regions implicated by the P-FIT relate to working memory, attention, language, and fluid reasoning, suggesting that integration of these cognitive processes is fundamental to intelligence.

Following publication of the P-FIT, ensuing research reliably corroborated the theory and in 2015 Basten, Hilger and Fiebach published a meta-analysis taking account of more recent methodological and technological improvements in imaging studies. The authors considered two types of studies: those examining neural correlates of experimental tasks, and those examining individual differences in brain activation and intelligence. Jung and Haier's review studies mostly used the group/task approach, while Basten et al. decided to investigate whether P-FIT regions related to individual differences in intelligence. Their findings corroborated the major premises of the P-FIT, and refined it. Basten et al. found less support for sensory region involvement than did Jung and Haier, which finding may be related to the individual differences rather than task-based focus of their meta-analysis. However, that the two papers, despite little overlap in the studies reviewed, reached broadly similar conclusions regarding the parieto-frontal intelligence network is noteworthy. The P-FIT remains the best-supported theory regarding the neural basis of intelligence.

Progressing the field of "neuro-metrics" will require clarification of its factor structures (e.g. Carroll, 1993); furthermore, to address previous methodological issues, such research will need to be grounded in an understanding of how individual differences in cognitive ability map onto differences in brain structure and function. Moreover, such measurements would need to discriminate reliably over the whole ability range – a persisting challenge for psychometrics at the extremes of ability. Particularly relevant to this study is how HIA brains differ to the general population, and the key differences at each level of neuroscientific analysis are summarised next.

*What are smart brains made of?*

While established that intelligence has a heritable component, locating "intelligence" genes is still an active line of research. A 2017 meta-analysis of GWAS studies encompassing 78,308 individuals (Sniekers et al.) uncovered 40 new loci, bringing the total number of genes implicated in intelligence to 52. Such findings, however, should be interpreted with caution: no gene codes for "intelligence" *per se*, but rather factors associated with brain development, e.g. the SHANK3 gene has been linked to synapse development (Zhubi et al., 2014). Furthermore, the genetic variants described by Sniekers et al. account for barely 5% of variance in intelligence; the so-called "missing heritability" problem in quantitative genetic research. Another open question is whether genes relating to HIA are qualitatively or quantitatively different from the population distribution (Shakeshaft et al., 2015). The discontinuity hypothesis proposes that HIA is aetiologically different, while the continuity hypothesis holds that HIA is genetically continuous with the normal intelligence distribution. Recent GWAS evidence (Zabaneh et al., 2017) from a sample of 1,238 individuals (mean IQ~170) supports the latter.

Structurally, there are modest but consistent differences in HIA brains. Intelligence correlates approximately .2 with head size and .4 with intracranial volume. Most papers reviewed by Luders et al. (2009) reported correlations around .3-.4 for global brain volume. Cortical convolution and intelligence were long believed to be related, but little quality research is available on either whole-brain or ROI convolutions (Luders et al., 2009). Cortical thickness studies reveal consistent differences, however, developmental trajectory is implicated more than cortical thickness alone. HIA has long been associated with cortical neuron arborisation (Ceci, 1990). A longitudinal study by Shaw et al. (2006) revealed HIA participants first display accelerated and prolonged cortical growth, followed by equally vigorous pruning around 8-12 years. VBM studies have shown around 6% of total grey matter volumes correlate with FSIQ, but with different location patterns according to age and sex (Haier et al., 2005). Colom et al. (2010) found intelligence correlated with grey matter approximately .31 and white matter approximately .27 (mainly frontal and temporal). Research has also found significant positive correlations between FSIQ and corpus callosum thickness (Luders et al., 2007). Modern DTI research also corroborates the brain myelination hypothesis of intelligence (Miller, 1994). Penke et al. (2012) investigated 12 white matter tracts that together explained 12% of variance in FSIQ, supporting the hypothesis that white matter integrity is related to speed and efficiency of cognitive processing.

Functionally, HIA brains appear to have different neurochemistry, blood flow, metabolic efficiency, and arousal patterns. One theory attributes human cognitive abilities to an expansion of the dopaminergic system (Previc, 1999), linked to regulation of motor planning, cognitive flexibility, temporal sequencing, imagination, working memory, and abstract reasoning. Cerebral blood flow relates to total neuron activity over time (Herscovitch, 1994), and CBF studies frequently report fewer changes in HIA participants (Haier et al., 1992). The earliest intelligence study using PET (Haier et al., 1988) revealed that, counterintuitively, participants with higher RAPM scores showed *less* neural activation during problem-solving, a finding that came to be known as the brain efficiency hypothesis. A simple linear relationship cannot be assumed, however. Studies have linked different patterns of arousal with different WAIS profiles, with high-arousal participants showing increases in some verbal index scores, and low-arousal participants showing increases in some performance scores (Robinson, 1989). Generally, however, HIA is linked with intermediate arousal levels.

Connectivity, neural activity and arousal are reflected in electrophysiological measurements. While mid-20$^{th}$ century EEG studies lacked technological and methodological sophistication, ERP analyses revealed brighter participants consistently showed earlier P300 responses and longer “string lengths” (wave complexity) than duller participants (Deary & Caryl, 1997). Such findings were, however, contradicted by Houlihan, Stelmack and Campbell (1998) who found longer P300 latencies in higher-ability participants. They proposed that a pure processing speed explanation was inadequate because brighter participants spent more time on analysis and planning, therefore shorter ERP latencies did not invariably signal higher IQ. Lee et al. (2012) used resting EEG to study the relationship of functional connectivity to the g-factor. They found relationships between posterior-anterior neural interaction in all frequency bands and drew two main conclusions: global connectivity strength predicted intelligence (but interestingly not processing speed); and regional functional properties also related to specific tasks. Langer et al. (2011) observed that resting EEG is a stable biomarker for intelligence-related neural activity, and proposed that since neurophysiological activations occur in milliseconds, EEG has clear advantages for studying network connectivity. Modern quantitative EEG techniques render efficient inspection of cortical and even subcortical activity (via source localisation software) making it simple to study with qEEG what was once only possible with fMRI.

*Measuring intelligence with EEG*

Thatcher and Lubar (2009) assembled a database of 332 qEEG recordings together with participants' educational, medical and psychometric data, including full-scale WISC/WAIS scores. Their multiple regression revealed numerous EEG variables correlating well with FSIQ. Three variables were particularly well-supported in the literature: individual alpha peak frequency (IAPF), power, and coherence.

Generation of the alpha rhythm has been traced to cyclic cortico-cortical and thalamocortical excitation patterns (Anokhin & Vogel, 1996). In most qEEG recordings there is a dominant peak of high-amplitude activity in the 7.5-12.5Hz range (Klimesch, 1999). This alpha peak reflects both state and trait levels of cortical arousal (Anokhin & Vogel, 1996), and research shows robust correlations between IAPF and FSIQ (Giannitrapani, 1985; Anokhin & Vogel, 1996). Cognitive deficits are associated with a low IAPF, while an IAPF over 10 signals high intelligence and/or anxiety. Provided overall levels of physiological arousal can be controlled for, it is hypothesised that higher-IQ individuals will display higher IAPF.

EEG power is the magnitude of each frequency band at each electrode and occurs due to the activity of synchronous populations of mostly pyramidal neurons over the whole cortex (Thatcher & Lubar, 2005). Giannitrapani (1985) conducted one of the most comprehensive early investigations into the relationship between IQ and EEG power and reported that verbal scores were positively related to alpha frequency power across all electrode sites. Jausovec (1996) found that HIA individuals showed higher alpha power in both resting and problem-solving conditions, and that mean alpha power across all electrodes was the power variable most significantly correlated with IQ. Power can be absolute (microvolts squared) or relative (percentage of power frequency X measured against the entire frequency spectrum). Absolute EEG power decreases with age (Klimesch, Vogt & Doppelmayr, 2000) but using relative power percentages instead serves as an age control. It is therefore hypothesised that higher-IQ individuals will display higher relative alpha power at all electrode sites.

Coherence estimates consistency of amplitude and phase between electrodes within a given frequency band (Bowyer, 2016), and determines how closely signals within different brain regions are correlated, and accords with the concept of functional connectivity within the brain. However, research has not been consistent in the direction of the relationship. According to both Thatcher and Lubar (2005) and Neubauer and Fink (2009), decreased coherence correlated with higher IQ. Furthermore, Langer et al. (2011) found higher IQ related to increased small-world network organisation in the upper alpha (10-12Hz) band. These findings support a model of focal specialisation and network complexity. Conversely, Yang and Chaou (2005) found higher intelligence to correlate with higher coherence, as did Lee et al. (2012), who interpret this finding as suggesting higher connectivity strengths in the fronto-parietal network. Although the direction of the correlation is unclear, it is predicted that there will be a relationship between IQ scores and frontal-parietal coherence in the 10-12Hz band.

While the EEG variables identified are predicted individually to relate to intelligence, they clearly do not function in isolation. It is therefore also hypothesised that there will be a relationship between IAPF, power, and coherence, and IQ scores.

### *Psychometrics and high IQ*

Spearman (1927) noted that correlations between psychometric tasks decreases in higher ability groups. Poor performers on one test type tend to perform poorly on them all, while higher ability individuals possess a more uneven ability profile. This “discontinuity hypothesis” was confirmed by Detterman and Daniel (1989), who divided the standardisation samples for the WISC-R and WAIS-R into five ability groups from low-high and analysed subtest correlations. Colom et al. (2002) subsequently found variance is better represented by group factors (e.g. verbal, performance) in higher ability populations. Jensen believes the g-factor breaks down altogether around four standard deviations from the population mean (Langan & LoSasso, 2002). In HIA ability populations it seems prudent therefore to report factor indices separately. Most factor analytic studies concur with a broad distinction between verbal and perceptual intelligence (Carroll, 1993). British Mensa offers two such tests for identifying very high IQ: the Cattell Culture Fair Intelligence Test (CFIT) and Cattell Verbal IIIB (CVIIIB). Use of the CFIT in recent research has also been identified (Duncan et al., 2000; Basten et al., 2015). The current study therefore selected the Cattell tests to measure verbal and perceptual reasoning, and seeks to extend previous research by examining IQ levels up to the 4SD+ range.

# METHODS

### *Design*

This study used a correlational and regressional design to explore possible relationships between the EEG variables of individual alpha peak frequency (IAPF), relative alpha power (percentage of 8-12Hz activity relative to the full EEG spectrum) across all electrode sites, and parietal-frontal coherence in the high alpha (10-12Hz) band, and verbal reasoning scores on the Cattell Verbal IIIB (CVIIIB) (Cattell & Cattell, 1960a) and perceptual reasoning scores on the Cattell Culture Fair Intelligence Test, Scale 3 (CFIT) (Cattell & Cattell, 1960b).

### *Participants*

Twenty-six participants (13 male, 13 female) were recruited variously from the University of Westminster student body, social media groups for Mensa, International Society for Philosophical Enquiry ("The Thousand") and other high IQ societies, a neuroscience-related Meetup group, the neighbourhood networking website Nextdoor.com, and miscellaneous personal contacts. Sampling was designed to incorporate all cognitive ability levels from low-average to exceptionally gifted. Test fees were refunded to those participants who took the Mensa-administered tests expressly to participate in the study. Because of difficulties recruiting participants in the average IQ range, the sample ended up heavily biased towards the high-ability end of the scale, with over half having IQs in the $98^{th}$ percentile or above. Some participants only had CVIIIB scores due to variations in Mensa test marking policy. The age of the participants ranged from 21 to 82 years. Two participants were left-handed. English language proficiency was native speaker (N=14) or academic-level fluency (N=12). Education ranged from secondary to doctoral level. Exclusion criteria included medication with deleterious cognitive effects, major head injury, or epilepsy. One participant did not provide his IQ scores and was therefore excluded from the study.

*Materials and Apparatus*

The CVIIIB and CFIT are timed, supervised, pencil-and-paper tests administered in a group setting. The CVIIIB is a mostly verbal-based multiple-choice test consisting of synonyms, classification, opposites, analogies, sentence completion, and inferences, while the CFIT is a perceptual reasoning multiple-choice test consisting of progressive series completion, classification (odd one out), matrix completion, and conditions (selecting the figure that most closely duplicates the pattern arrangement of the target stimulus). The CVIIIB has a ceiling score of 161(SD24), while the CFIT has a ceiling score of 183(SD16), making the latter especially suitable for identifying very high cognitive ability. To make comparisons between the two tests simpler, all IQ scores in this study have been converted to the modern SD15 scale.

The EEG system used was a Mitsar 202 portable amplifier (Mitsar Co. Ltd., St. Petersburg, Russia), attached to an ECI (Electro-Cap International, Eaton, OH) EEG cap with inbuilt tin electrodes arranged according to the International 10/20 system, and also attached were separate ECG and respiration sensors. WinEEG version 2.84.44 (Mitsar Co. Ltd.) software was used to record and save all the electrophysiological data.

PsyTask version 1.53.17 (Mitsar Co. Ltd.) was used to administer a visual continuous performance task (a modified Go-NoGo task). This task presents 100 random trials of four possible pairings of stimuli: animal-animal (the Go trial), animal-plant (the NoGo trial), plant-plant (the Ignore trial) and plant-human combined with one of five possible pure tone sounds (the novel stimulus trial). Stimuli are presented for 100ms each, the interval within pairings is 1100ms and the inter-stimulus interval between trials is 3100ms. Participants must click a mouse as rapidly as possible only when presented with the animal-animal pairing and refrain from clicking on any of the other trials. Both event-related potentials (ERPs) and behavioural data are recorded during this task.

*Procedures*

*Cognitive Assessment*

Participants took the CVIIIB and CFIT at their local British Mensa test centre on a date of their choice, some of which were before and some after the EEG session. Testing was invigilated and scored by a qualified examiner. Participants received their test scores by mail within 10 days of testing from British Mensa Head Office and the author had sight of the test report letters.

*Electrophysiological Recordings*

Before the EEG session participants were instructed to get a good night's sleep, to wash their hair and not use hair products (e.g. gel) since these can interfere with electrode connectivity to the scalp, not to consume caffeine after breakfast time, and to avoid wearing contact lenses during the recording as they increase blinking.

Nineteen-channel quantitative EEGs were recorded (electrodes Fp1, Fp2, F7, F3, Fz, F4, F8, T3, C3, Cz, C4, T4, T5, P3, Pz, P4, T6, O1 and O2) using common average montage, linked ears referential, and electrode anterior to Fz as the ground. Impedance was maintained below 5kOhm. Sampling rate was 2,000Hz, down-sampled to 500Hz when saved. ECG and respiration data were recorded via sensors on the wrists and a monitor waistband respectively. Five minutes' EEG was recorded with eyes closed (EC) and five minutes with eyes open (EO), with a short break at each halfway point to ensure participants remained focused and comfortable. Twenty minutes of ERP data was also recorded (in two 10-minute blocks) while participants performed the visual continuous performance (VCPT) task. The ERP data are not analysed in the current study but the behavioural data from the task (errors of omission and commission, reaction times, and reaction time variability) are examined.

*Data Pre-processing*

EEG data was de-artefacted in the WinEEG software using Independent Components Analysis and automatic artefact rejection. Coherence values between all pairs of electrodes in all bandwidths (delta, theta, alpha1, alpha2, beta1, beta2 and gamma) were computed in WinEEG using the coherence analysis function. Relative EEG power in each frequency band at each electrode site was computed using the WinEEG spectral analysis tool. Finally, each participant's alpha peak frequency was identified by locating the maximal difference peak (usually occipital or parietal) during alpha blocking (EC minus EO). Data were exported from WinEEG via text files to an Excel spreadsheet, where the exact variables of interest were identified for analysis. HRV data was computed from the heart rate and respiration data recordings using custom MATLAB scripts.

*Data Analysis Strategy*

The three EEG variables in the current study were chosen because of their moderate to relationship with IQ (Thatcher & Lubar, 2009). The IAPF-IQ relationship was well supported in the literature (e.g. Anokhin & Vogel, 1996; Doppelmayr et al., 2002; Klimesch, 1999; Polunina & Davydov, 2006), as was that of relative alpha power (e.g. Jausovec, 1996; Langer et al., 2011; Polunina & Davydov, 2006; Thatcher & Lubar, 2005). HRV data was also collated to control for overall physiological arousal. Power was averaged over all electrode sites, as per Jausovec and Jausovec (2000). Since alpha power changes across the lifespan (Klimesch et al., 2000), relative rather than absolute alpha power was selected to control for the varying ages of participants. Coherence data in the frequency band alpha2 (10-12Hz) was selected because activity in this bandwidth had been identified as correlating most closely with IQ scores (e.g. Klimesch, 1999; Langer et al., 2012). Furthermore, Thatcher and Lubar (2009) found coherence to be one of two variables strongly correlated with IQ (r=.84). Rather than analyse all 171 electrode pairs, electrodes Fp1/P3 and Fp2/P4 were selected because Thatcher et al. (2016) found the largest differences between high and low IQ in phase slope index (a neural connectivity variable closely linked to coherence) between these electrode sites in the alpha frequency bands. These sites also match the locations of the Brodmann areas identified as relevant to the P-FIT (Jung & Haier, 2007). It only remained to be determined whether focusing on electrodes over the parietal-frontal network specifically would reveal patterns of activity more pronounced in very high-performance brains.

## *Ethical Issues*

This study received consent from University of Westminster Ethics Committee. No vulnerable populations were involved and there were no potentially harmful procedures. Participants were informed regarding the purpose and procedures of the study, and written consent was obtained. Debriefing and opportunities to ask questions were provided. Participants understood that they could withdraw from the study at any time. Data collated for analysis were kept confidential and fully anonymised.

# RESULTS

*Cognitive, Demographic & Behavioural Data*

On the two Cattell tests, the mean CVIIIB score was 127.04 (SD 19.26) (min. 69, max. 160), and the mean CFIT score was 124.05 (SD 23.61) (min. 82, max. 178), both of which sets of scores were well above the population mean IQ of 100. As expected, scores on the two psychometric tests correlated highly with each other ($r$=.75, $p$=<.01, one-tailed).

Preliminary exploration of the psychometric test scores against participants' demographic data revealed that CVIIIB scores were positively correlated with age ($r$=.42, $p$=.020, one-tailed) which was as expected, but interestingly so were CFIT scores ($r$=.51, $p$=.009, one-tailed) even though fluid intelligence generally declines with age (Horn, 1982). There were also no significant differences between males and females on either CVIIIB or CFIT scores (see **Appendix 4**). Furthermore, there was no significant difference between native and non-native speakers of English with regard to CVIIIB scores (see **Appendix 4**) although non-native speakers had a greater spread of scores on this test than native speakers. However, unexpectedly, non-native English speakers scored significantly lower on the CFIT: 108.73 compared to 140.90 ($t$=4.24, $df$=19, $p$=<.01, two-tailed). Education level is often used in research as a proxy for IQ (Deary & Johnson, 2010), however, in this sample education level had no bearing on either CVIIIB or CFIT scores see **Appendix 4**).

On the 20-minute visual continuous performance task, most participants made zero or only one error of omission (missing a trial) or commission (responding in error) (see **Appendix 3**), with an average reaction time of 375.16ms, and an average reaction time variability of 6.85ms. Most participants performed more accurately, faster, and more consistently on this task than the norms for their age group, while many performed the task better than every age group for which norms were available. The accuracy scores in particular showed a ceiling effect.

*EEG & Physiological Data*

The EEG variables of interest were explored as follows. The IAPF of the sample was 9.96Hz (SD 0.84), which is within the usual range of 7.5-12.5Hz (Klimesch, 1999), although two participants had the “low amplitude fast” EEG phenotype (Gunkelman, 2001) without any discrete alpha peak. Participants’ mean EEG alpha power was 18.82% relative to the full frequency spectrum, but individual values were quite disparate (SD 10.58%) across the sample. The mean coherence values for the sample (the software generates coherence scores as a correlation coefficient) are shown in **Table 1** below for the two pairs of electrodes selected, in both eyes closed and eyes open conditions.

**Table 1: EEG Coherence Values (High Alpha band: 10-12Hz)**

| **Electrode Pair** | **Condition** | **Coherence (standard deviation)** |
|---|---|---|
| Fp1-P3 | Eyes closed | 0.41 (0.18) |
| Fp2-P4 | Eyes closed | 0.42 (0.22) |
| Fp1-P3 | Eyes open | 0.35 (0.20) |
| Fp2-P4 | Eyes open | 0.37 (0.21) |

The three EEG variables and their correlations with the two IQ scores are reported in **Table 2** below. Correlations between IAPF and relative alpha power and the two Cattell tests did not show any significant relationship, and the experimental hypotheses that IAPF and relative alpha power would be positively related to the two IQ scores was rejected. There was only one significant result for coherence, between electrodes Fp1 and P3 in the eyes-closed condition and scores on the CFIT, but since corrections for multiple comparisons were not run it is unclear whether this was a chance result. Given the proportion of other nonsignificant findings the experimental hypothesis that there would be a relationship between alpha coherence and the two IQ scores was rejected.

**Table 2: Correlations between Individual Alpha Peak Frequency, Relative Alpha Power & Coherence and Cattell Test Scores**

| EEG Variable | IQ Test | Correlation | Result |
|---|---|---|---|
| Alpha Peak Frequency | CVIIIB | *r*=.10, *p*=.329, one-tailed | Not significant |
| | CFIT | *r*=.02, *p*=.471, one-tailed | Not significant |
| Alpha Power | CVIIIB | *r*=.18, *p*=.206, one-tailed | Not significant |
| | CFIT | *r*=-.25, *p*=.139, one-tailed | Not significant |
| Coherence | | *(Two tailed tests used as direction in previous literature contradictory.)* | |
| Fp1-P3 EC | CVIIIB | *r*=.15, *p*=.491 | Not significant |
| | CFIT | ***r*=.49, *p*=.024** | **Significant** |
| Fp2-P4 EC | CVIIIB | *r*=-.18, *p*=.384 | Not significant |
| | CFIT | *r*=-.06, *p*=.781 | Not significant |
| Fp1-P3 EO | CVIIIB | *r*=.27, *p*=.191 | Not significant |
| | CFIT | *r*=-.07, *p*=.765 | Not significant |
| Fp2-P4 EO | CVIIIB | *r*=.09, *p*=.669 | Not significant |
| | CFIT | *r*=-.08, *p*=.718 | Not significant |

Partial correlations were also run on IAPF scores with the two Cattell tests, and controlling for heart rate variability scores (as an overall index of physiological arousal/anxiety). Overall, the sample had generally healthy heart rate variability (71.35ms, SD 11.14ms), and partialling out the HRV scores had a negligible effect on the correlations reported above (see **Appendix 5**). No significant relationship was found between IAPF and either Cattell test after controlling for overall arousal levels.

Although few correlations had thus far reached significance, regressions were nevertheless performed to explore whether the three EEG variables (IAPF, relative alpha power, and coherence) would cumulatively have any bearing on Cattell test scores. In the case of coherence, the value having the best correlation with IQ (Fp1-P3, eyes closed), was selected. The three EEG variables were entered as predictors into separate multiple regressions for the CVIIIB and CFIT using the standard method. However, no significant model emerged for either the CVIIIB ($F$(3,19)=.73, $p$=.544) or CFIT ($F$(3,16)=1.73, $p$=.201) from these predictor variables (see **Appendix 5**).

# DISCUSSION

This study explored possible correlations between the EEG variables of individual alpha peak frequency (IAPF), relative alpha power (8-12Hz, averaged across electrodes), and 10-12Hz alpha coherence between electrodes Fp1-P3 and Fp2-P4 in eyes closed and eyes open conditions, with scores on a measure of verbal reasoning (Cattell Verbal IIIB) and a measure of perceptual reasoning (Cattell Culture Fair), and also explored possible effects of all three EEG variables on each of the two Cattell test scores.

No significant relationship was found between IAPF and either Cattell test. This finding did not replicate previous research (e.g. Anokhin & Vogel, 1996; Doppelmayr et al., 2002; Klimesch, 1999; Polunina & Davydov, 2006), which found a positive relationship between IAPF and IQ. All IAPF values were within the expected range of 7.5-12.5Hz (Klimesch, 1999), although two participants had an EEG profile that fit the “low amplitude fast” phenotype, a normal variant pattern dominated by low amplitude beta-wave activity with no discernible alpha peak (Gunkelman, 2001). Heart rate variability scores were within expected ranges for fit, healthy adults, where higher scores indicate higher parasympathetic activity and low scores can be indicative of anxiety and stress. Partialling out HRV scores to control for physiological arousal (as a proxy for trait anxiety) had a negligible effect on IAPF/IQ correlations, suggesting that participants’ physiological arousal levels were fairly similar across the group.

There was also no significant relationship between either Cattell test and relative alpha power, which again did not replicate previous research (e.g. Jausovec, 1996; Langer et al., 2011; Polunina & Davydov, 2006; Thatcher & Lubar, 2005) where a relationship between relative alpha power and IQ had been identified. This study examined combined alpha from the alpha1 (8-10Hz) and alpha2 (10-12Hz) bandwidths; a more fine-grained analysis of the individual EEG frequencies may render different results. Furthermore, relative power was used to control for the wide age range of the participants, since alpha power changes over the lifespan (Klimesch et al., 2000). However, Thatcher and Lubar (2005) found smaller effect sizes of relative power than absolute power when looking for EEG/IQ correlations. However, participants of similar age would need to be recruited if using absolute EEG power as the power variable.

Furthermore, no significant relationships were found between either Cattell test and 10-12Hz alpha coherence, except for one positive correlation between CFIT scores and coherence at electrodes Fp1-P3 in the eyes-closed condition. This single positive relationship contradicted previous research that reported negative correlations between coherence and IQ (Neubauer & Fink, 2009; Thatcher & Lubar, 2005) but supported the findings of Yang and Chaou (2005) and Lee et al. (2012) who reported positive correlations. It is difficult to interpret this single result given the number of other nonsignificant findings, but Jung and Haier (2007) reported greater left hemispheric activation for some P-FIT areas so, if meaningful, this finding would imply greater parietal-frontal left-hemisphere upper alpha connectivity is related to higher perceptual reasoning ability. Since most of the findings were very clearly not significant, no post-hoc correction for multiple comparisons was carried out as it seemed unlikely such would alter the results, but it remains possible that this single positive correlation was a chance finding. Again, a more refined analysis of single EEG frequencies may render a different result.

Multiple regressions to explore possible correlations between all three EEG variables and, separately, the two Cattell tests did not find any significant relationships, a finding which did not replicate the multiple regression by Thatcher and Lubar (2005) which identified some dozen or so EEG variables modestly to strongly correlated with intelligence. Thatcher and Lubar, however, used a very large number of predictor variables that they had also checked using discriminant analyses between high and low IQ samples and there may have been large differences in statistical analysis techniques between Thatcher and Lubar and the current study.

In view of the foregoing, the experimental hypotheses that relationships would exist between the EEG variables and verbal and perceptual reasoning scores were not supported by the data and were rejected. The lack of replication of previous findings, however, may have been because this study differed in several ways (participants, psychometric tests, and design) that cumulatively may have had an effect on the overall findings.

Because of the small number of available participants, it was not possible to select for handedness or sex; however, score differences between the sexes were nonsignificant. On the other hand, it was possible to control for age, in the EEG data by using relative rather than absolute alpha power (Klimesch et al., 2000), while the Cattell tests are age-normed (Cattell & Cattell, 1960a; 1960b). It is also possible that the nonsignificant findings are linked to the very atypical mean IQ scores of the participants. Because of recruitment difficulties in the average IQ range, more than half the participants had IQs >2SD above the population mean. The findings therefore cannot be generalised to the wider population.

Detterman and Daniel (1989) established that psychometric variables correlate less well in HIA populations and render a "spikier" cognitive profile. Furthermore, it is known that HIA individuals tend to score lower on processing speed relative to other cognitive domains (Watkins, Greenwall & Marcell, 2002). This finding implies that speed no longer relates to cognitive "power" beyond a certain IQ level and, if included in a battery composite, may drag FSIQ down. The Cattell tests are administered under very tightly-timed conditions, and it has been suggested that highly-speeded test administration may introduce construct-irrelevant sources of variance (Lu & Sireci, 2007). The irrelevance of speed in an above-average IQ group was supported by a notable ceiling effect on the VCPT task, with most participants making few or zero mistakes, and demonstrating reaction times and reaction time variability faster than the norms for their age group.

Previous research also typically used either a Wechsler battery (to derive FSIQ) or Raven's Matrices (as a proxy for g). The current study, on the other hand, used tests of verbal and perceptual reasoning, rather than a composite IQ score. Continued research emphasis on a unitary structure for intelligence has meant that little neuroscientific investigation has been conducted into first and second order CHC theory factors (Stankov, 2017). To identify the neural underpinnings of "intelligence in general" rather than "general intelligence" (Colom et al., 2002), these lower order factors will need to be investigated thoroughly. To date, there has been too little functional imaging work using HIA participants to fully clarify the neural mechanisms of very high IQ. Specific imaging and/or EEG work in HIA populations will be required, along with the use of a more sophisticated cognitive test, preferably a full battery with a low speed loading, with extended norms for valid measurements in the high range.

In terms of design, most literature reviewed for this study looked for differences between groups of participants assigned by IQ level. It may have been easier to obtain a significant result comparing groups with smaller numbers of participants, whereas this study used a correlational and regressional design which, ideally, would have benefited from a larger sample size.

Given the foregoing, therefore, the lack of significant findings in the current study should be treated with caution and not necessarily be interpreted as contradicting any previous EEG and imaging studies or existing theory of intelligence. It is proposed that, just as psychometric g weakens and eventually unravels completely in HIA populations, so that high psychometric performance relies increasingly on specialised abilities (Spearman, 1927), if there is such construct as a “neuro-g”, the neural processes giving rise to it also cease to correlate well with higher intelligence, with high performance brains becoming increasingly neurologically specialised. Haier (2009) pointed out that psychometric g may not correspond in a unitary fashion with brain structure. Certainly, the neural processes underpinning individual WAIS task performance remain poorly understood (Polunina & Davydov, 2006). However, research to date suggests that the neural underpinnings of intelligence map somewhat better onto two dimensions (Nisbett et al., 2012) rather than one. The strongest support so far has been for the distinction between fluid and crystallised intelligence (Horn & McArdle, 2007), hence building on the current study’s exploration of verbal and perceptual abilities seems a reasonable starting point.

Other variables were obtained during data collection that could not be analysed in the current study due to their number and complexity, and it is possible that a more comprehensive evaluation of the recorded data may more fully elucidate the relationships between the EEG measures and IQ scores. Event-related potentials recorded during the VCPT task were not analysed and these data may reveal greater relationships with the IQ scores than the resting EEG alone. Although 10-12Hz alpha was the waveband most strongly correlated with FSIQ or g in the literature, it remains an open question how different EEG frequencies correlate with first- and second-order CHC factors. This study examined coherence patterns at four electrode sites (Fp1-P3 and Fp2-P4); it was not feasible to investigate all 171 possible pairings. There is a suggestion that different patterns of activation occur between short-range networks and long-range connections (Thatcher et al., 2016). Furthermore, that Jung and Haier’s 2007 review also suggested some P-FIT regions activated bilaterally while others showed more prominent left hemispheric activation merits additional follow-up. Future research should focus on addressing the methodological limitations of the current study previously described, and also exploring the additional variables that were identified but could not be analysed in this study.

Given the substantial work still needed to connect psychometric measures with their neural foundations, it is unlikely that “neuro-metrics” will replace psychometrics in the foreseeable future. The field of “differential neuroscience” (Deary, 2012) is still very much in its infancy. It is envisaged that as this psychometric/neuroimaging research progresses, a new paradigm will eventually emerge connecting the CHC theory expounded by Carroll (1993) and Jung and Haier’s (2007) P-FIT model, which will serve as the basis of future experimental imaging work in the field of intelligence. However, any new theoretical intelligence framework will need to possess sufficient predictive validity and reliability to measure all the way to the extremes of the ability scale to be useful in furthering our understanding of very high intelligence.

To summarise, this study sought to explore the relationships between the EEG variables of individual alpha peak frequency, relative alpha power, and alpha coherence, and measures of verbal and perceptual reasoning. The findings generally were not significant and did not accord with earlier research findings on EEG and IQ. However, no previous EEG study included such a large proportion of high intellectual ability participants, who may have considerably different psychometric and neurological profiles to the majority of the population and the findings cannot be generalised. Furthermore, this study highlighted many EEG and cognitive variables that remain thus far unexamined. Finally, attention is drawn to the need for future intelligence research to work towards connecting existing psychometric and neuroscientific paradigms.

(6,907 words)

# APPENDICES

Appendix 1: Ethics Approval
Appendix 2: Participant Information Sheet, Consent & Debrief Sheet
Appendix 3: SPSS Output – Descriptive Statistics
Appendix 4: SPSS Output – Initial exploration of the data
Appendix 5: SPSS Output – EEG/IQ Correlations and Regression